\documentstyle[twoside,fleqn,espcrc2,epsf,epsfig]{article}
\newcommand{\AmS}{{\protect\the\textfont2
  A\kern-.1667em\lower.5ex\hbox{M}\kern-.125emS}}
\newcommand{\gsim}{\mathrel{\mathop{\kern 0pt \rlap
  {\raise.2ex\hbox{$>$}}}
  \lower.9ex\hbox{\kern-.190em $\sim$}}}  

\title{\Large{\bf ROM2F/97/39 - october 24, 1997 - contributed paper 
to TAUP97 - to appear in the Proceedings \\
-------------------------------------------------------}\\ 
WIMPs search by scintillators: possible strategy for annual 
modulation search with large-mass highly-radiopure NaI(Tl)}

\author{R. Bernabei$^a$, P. Belli$^a$, F. Montecchia$^a$,
W. Di Nicolantonio$^b$, A. Incicchitti$^b$, D. Prosperi$^b$, 
C. Bacci$^c$, 
C. J. Dai$^d$,L.K. Ding$^d$,  
H.H. Kuang$^d$, J.M. Ma$^d$  \\
{\it $^a$ Dipartimento di Fisica,
II Universit\`a di Roma  and
I.N.F.N. Sezione di Roma2, Italy;
$^b$ Dipartimento di Fisica,
Universit\`a di Roma 
and I.N.F.N. Sezione di Roma, Italy.
$^c$ Dipartimento di Fisica,
III Universit\`a di Roma 
and I.N.F.N. Sezione di Roma, Italy;
$^d$ IHEP, Chinese Academy, P.O. Box 918/3, Beijing 100039,
China.}}

\begin{document}

\begin{abstract}
The DAMA experiments are running deep underground in 
the Gran Sasso National Laboratory. Several interesting
results have been achieved so far. Here a maximum likelihood
method to search for the WIMP annual 
modulation signature
is discussed and applied to a set of preliminary test data 
collected with large mass highly radiopure NaI(Tl) detectors.
Various related technical arguments are briefly addressed.

\end{abstract}

\maketitle

\section{Introduction.}

   The DAMA experiments 
have already achieved several interesting
results using various target-detectors \cite{Ber96,Bel96,Dam97}. 
This paper is devoted to discuss an analysis strategy
on annual modulation signature; a practical example
is performed by using a preliminary data set 
collected with large-mass highly
radiopure NaI(Tl) detectors. Standard assumptions 
for the WIMP model have been considered.
A more extensive
discussion on analysis strategies is in \cite{Ber97}. 
 
  In the direct searches by WIMP-nucleus elastic scattering,
the possible presence of a WIMP signal 
can be extracted from the background by the 
annual modulation of the WIMP rate
in the target-detector \cite{Bel96,Dru86,Fre88,Sar96}.  
In fact, the expected recoil 
energy spectrum depends on the WIMP velocity distribution 
and on the Earth velocity in the galactic frame, v$_r$(t),
which varies along the year according to the expression:
$v_r(t) = V_{Sun} + V_{Earth} cos\gamma cos \omega(t-t_0)$.
Here V$_{Sun}$ = 232 km/s is the Sun velocity with respect 
to the galactic halo and V$_{Earth}$ = 30 km/s is the Earth orbital 
velocity around the Sun on a plane with inclination 
$\gamma$ = 60$^0$ with respect to the galactic one; 
furthermore, $\omega$ = 2$\pi$/T with 
T=1 year and t$_0 \simeq 2^{nd}$ June (when the Earth speed 
is at maximum).  The WIMP velocity distribution in 
the galactic halo frame is considered to be a Maxwellian 
distribution with v$_0$ parameter (defined as 
$\sqrt{2/3} \cdot v_{r.m.s.}$) and a cut-off velocity 
equal to the escape velocity. The Earth velocity can be 
conveniently expressed in unit of
v$_0$ : $\eta$(t) = v$_r$(t)/v$_0$  =  $\eta_0$ + $\Delta\eta$ cos$ \omega(t-t_0)$, 
being $\eta_0$ =1.05 the yearly
average of  $\eta$ and $\Delta\eta$ = 0.07. 
Since $\Delta\eta << \eta_0$, 
the signal rate in the k-th energy interval is accurately 
given by the first order Taylor approximation:
$ S_k[\eta(t)] = S_k[\eta_0] + [{\delta S_k \over 
\delta\eta}]_{\eta_0} \Delta\eta cos\omega(t-t_0) 
= S_{0,k} + S_{m,k}cos \omega(t-t_0)$.
The contribution from the highest order terms is lower than 0.1\%. 
To select the S$_{m,k}$ the highest sensitivity is obtained when 
considering the smallest energy bins allowed by the available 
statistics and whole year data taking \cite{Ber97}.
  Although the time dependent effect is expected to be small, a suitable 
large-mass low-radioactive set-up with an efficient stability
monitoring can point out its presence
by time correlation analysis \cite{Ber97,Fre88}. 
Note that, with the present 
technology, the annual modulation remains the main signature of
a possible WIMP signal\footnote{We comment, in particular, 
that a pulse shape discrimination --- even under the 
assumption of an "ideal" electromagnetic background 
rejection --- cannot account alone for a WIMP signature. 
In fact, e.g. the neutrons and the internal end-range $\alpha$'s  
induce signals indistinguishable from WIMP induced 
recoils and cannot be estimated and subtracted in any
reliable manner at the needed precision.}.

\section{The experimental data}
  The data considered here have been collected with nine
9.70 kg NaI(Tl) detectors, part of the 115.5 kg highly
radiopure NaI(Tl) set-up now running at the Gran
Sasso Laboratory \cite{Ber97}. A description of the detectors
considered here and of the shield can be found in \cite{Ber96}.
We recall only that each
detector is viewed by two low-background EMI photomultipliers
working in coincidence at single photoelectron threshold;
2 keV is the software
energy threshold. It corresponds in our case to
$\gsim$ 11 photoelectrons (depending on the detector),
i.e. well distinguishable from PMT noise. In fact, this
last is present as fast, single, spare photoelectrons,
while the "physical" pulses have a time distribution
with a decay time of hundreds ns\footnote{We recall 
in addition that 
scintillators are not affected by microphonic noise as
ionizing and bolometer detectors.}.

\begin{figure}[h]
\centerline{\epsfig{figure=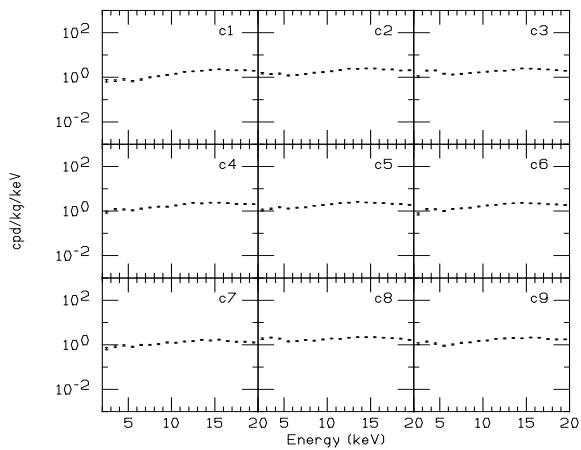,angle=90,height=4.0truecm,width=4.5truecm}}
\caption{Measured energy distributions for nine detectors.}
\end{figure} 

   A statistics of 4549.0 kgxdays is available for the
annual modulation studies: 3363.8 kgxdays in winter
time and 1185.2 kgxdays in summer period. It is
distributed along the year in the following way:
$-1 \le   cos\omega(t_i-t_0) \le -0.334$ in winter time and
$0.932 \le  cos\omega(t_i-t_0) \le 0.996$ in summer time.
It is evident that part of the time was devoted 
to systematic studies on calibrations and detector 
features and qualification (more extensively in the winter 
period, being the starting one), that is
all the available low energy data 
collected at that time have been considered. 
Obviously the analysis method properly takes 
into account both the collected low energy statistics 
and the cosine range in which it has been taken.
In fig.1 the measured energy distributions are shown.     

 The stability control can profit of the $^{210}$Pb peak at
46.5 keV present at level of a few cpd/kg, mainly
due to a surface contamination by environmental
radon during the first period of crystals storage
underground; both the peak position and the resolution
are controlled binning together the data
every $\simeq$ 7 days (see fig.2). This allows an intrinsic
monitoring of the threshold  and of the calibration stability
(e.g. PMT gain and electronic line stability).

\begin{figure}[h]
\centerline{\epsfig{figure=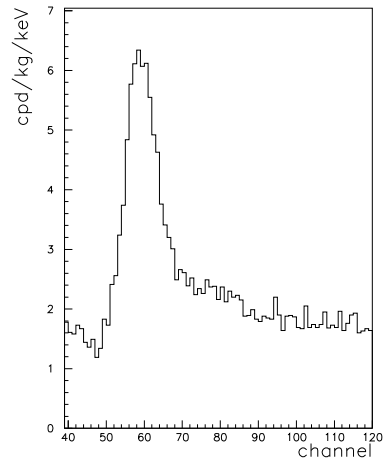,height=4.5truecm,width=7.0truecm}}
\caption{Typical $^{210}$Pb peak at 46.5 keV (see text).}
\end{figure} 

 Note that here we do not consider
any pulse shape analysis (PSA) to reject electromagnetic 
background (see \cite{Ber96,Smi96}).
In fact: i) the PSA has --- in any case ---
a statistical nature; ii) whenever it gives upper limits on the 
signal (as it is generally the case), its results cannot be used 
for the annual modulation analysis; iii)  the annual modulation
analysis acts itself as a very efficient background rejection. 
In any case, as shown in fig.3, the result obtained here
is well compatible with the upper limits obtained 
by using PSA considering also these detectors 
and published in \cite{Ber96}.

\section{The maximum likelihood method and the quest for a candidate}

  To determine the cross section and mass of a possible 
WIMP candidate, a time correlation analysis of all the 
data --- properly considering  the time 
occurrence and energy of each event --- has been performed.  

  The experimental data collected by a j-th detector of mass 
M$_j$ are considered as grouped in 1-day time bins and 
in $\Delta$E = 1keV energy bins; the number of the events in the
 i-th day and k-th energy bin, N$_{ijk}$, will follow a poissonian 
statistics with mean value given by
$\mu_{ijk} = (b_{jk} + S_{0,k} + S_{m,k} cos\omega(t_i-t_0)) M_j \Delta t_i \Delta E \epsilon_{jk},$
where a time independent background contribution, b$_{jk}$, 
has been included in addition to the dark matter signal searched for.
Here $\Delta t_i$ represents the detector running time during the i-th
day ($\Delta t \le $ 1 day) and $\epsilon_{jk}$ 
represents the analysis cut efficiency
\cite{Ber96}. The maximum likelihood function 
results: 
$  {\it\bf L}  = {\bf \Pi}_{ijk} e^{-\mu_{ijk}} {\mu_{ijk}^{N_{ijk}} \over N_{ijk}!}$.

Firstly, to allow a direct simple comparison 
of the sensitivity reached here with those of
previous experiments \cite{Bel96,Sar96}, 
the function 
$y = - 2ln({\it\bf L}) - const = \Sigma_ky_k$, 
can be minimized with respect to (b$_{jk}$+S$_{0,k}$) and 
S$_{m,k}$, 
using for each k-th
energy bin the data collected at the same time
with the 9 crystals; the $y_k$ 
can be here minimized independently.
The obtained S$_{m,k}$ values are shown in table 1; 
they point out the relevance of 
increasing the collected statistics.

\begin{table}[!htbp]
\begin{center}
\caption{S$_m$ values obtained by the maximum likelihood method.}
\begin{tabular}{|c|c|||c|c|}

\hline
\hline

 Energy &  S$_m$  & Energy &  S$_m$ \\
 (keV)  & (cpd/kg/keV) & (keV)  & (cpd/kg/keV) \\ \hline
 \hline
2-3    & 0.023 $\pm$ 0.037 & 11-12  & 0.053 $\pm$ 0.027 \\ \hline 
3-4    & 0.017 $\pm$ 0.030 & 12-13  & 0.015 $\pm$ 0.028 \\ \hline 
4-5    & 0.036 $\pm$ 0.027 & 13-14  & 0.017 $\pm$ 0.029 \\ \hline  
5-6    & 0.042 $\pm$ 0.021 & 14-15  & 0.023 $\pm$ 0.030 \\ \hline
6-7    & 0.038 $\pm$ 0.022 & 15-16  & -0.045 $\pm$ 0.029 \\ \hline
7-8    & 0.003 $\pm$ 0.023 & 16-17  & 0.008 $\pm$ 0.030 \\ \hline 
8-9    & 0.050 $\pm$ 0.024 & 17-18  & -0.031 $\pm$ 0.029 \\ \hline
9-10   & 0.065 $\pm$ 0.025 & 18-19  & 0.002 $\pm$ 0.031 \\ \hline 
10-11  & 0.032 $\pm$ 0.026 & 19-20  & 0.016 $\pm$ 0.030 \\ \hline
\end{tabular}
\end{center}
\end{table}      

However, from this table, a first qualitative 
view can be also obtained. In fact, e.g. in the first part of the measured 
energy region, 2-12 keV 
(where a signal contribution can {\it "a priori"} be expected),
$<S_m>_{2-12 } = (0.037 \pm 0.008)$ cpd/kg/keV 
\footnote{This first qualitative 
approach is introduced in \cite{Ber97} during 
a preliminary discussion on the Freese et al. method; the values were 
quite similar (see \cite{Ber97} for details).}
is obtained, while 
in the second one, 12-20 keV (where any signal 
is {\it "a priori"} expected to be suppressed),
$<S_m>_{12-20}  = (0.000 \pm 0.010)$ cpd/kg/keV is found.
The first value qualitatively supports the presence of a 
possible modulation, requiring therefore a suitable 
statistical analysis of the data to verify if it can be in total or in part 
(as quantified later by the C.L. and the $\chi^2$ values)
ascribed to a possible WIMP presence, or it can not.
The second value well supports
the absence of an overall systematics significantly 
exceeding the statistical error in the interesting energy 
region. 

We stress 
that here and in \cite{Ber97} 
the achieved result (see later) is not linked to this qualitative 
description (that is, it is not linked to any $<S_m>_{2-12}$ value), 
but to the 
application of the maximum likelihood method {\it on all
the available data (2-20 keV)} 
described in the following. Also an unbiased
statistical analysis on all 
the available data at the same time is then performed to obtain 
quantitative estimates on a possible effect. 

The maximum likelihood method
is well suitable to test the following hypothesis: 
the modulated effect qualitatively
described above can (or not)
be accounted by a WIMP with cross section and mass 
allowed for istance for the neutralino \cite{Bot97}, assuming a 
Spin-Independent (SI) interaction and, if it can, at which C.L..  
In particular, we 
refer our results to the quantity $\xi\sigma_p$ with 
$\xi = {\rho_{WIMP} \over \rho_0}$,
$\rho_0$ = 0.3 GeVcm$^{-3}$ and $\sigma_p$ WIMP cross section 
on proton. For this purpose, we note that 
S$_{0,k}$ and S$_{m,k}$ can be written ---  pointing out 
their dependence on $\xi\sigma_p$ and the WIMP mass, M$_w$  --- in the form 
$S_{0,k} = \xi\sigma_p S\prime_{0,k}(M_w)$ 
and  $S_{m,k} = \xi\sigma_pS\prime_{m,k}(M_w)$ 
respectively. The $S\prime_{0,k}(M_w)$ and $S\prime_{m,k}(M_w)$ can be
calculated according to \cite{Ber96} (e.g. besselian form
factor, $\xi$=1, etc.). The $\xi\sigma_p$ and M$_w$ values in the 
best agreement with the experimental data have
been obtained by minimizing here the $y$ function ---
using all the events with
their energy and time occurrence  ---
with respect to the free parameters: $\xi\sigma_p$,  M$_w$ and b$_{jk}'$s. 
A two-step 
minimization strategy allowed us to handle this large number
of parameters. In fact, by a preliminary y minimization the 
(b$_{jk}$+S$_{0,k}$)=f$_{jk}$ values have been determined and then, 
by a subsequent one, also the $\xi\sigma_p$ and M$_w$  values. 
In this last step, the conditions $M_w \ge $25 GeV 
and  (b$_{jk}$+S$_{0,k}$) = f$_{jk}$ if 
$\xi\sigma_p \le  {f_{jk} \over S\prime_{0,k}}$  
or (b$_{jk}$+S$_{0,k}$) =$\xi\sigma_p S\prime_{0,k}$ 
otherwise are required, to take into account both the results 
achieved at accelerators for SUSY particles and the obtained 
values for the unmodulated term.

  The minimum value of the y function has been found
for M$_w$ = (59\tiny \( \begin{array}{l}+36\\-19\end{array}\)\normalsize)
GeV and $\xi\sigma_p$ = 
(1.0\tiny \( \begin{array}{l}+0.1\\-0.4\end{array}\)\normalsize) 
10$^{-5}$ pb \cite{Ber97}.

\section{Consistency checks and statistical evaluations}

  To verify the consistency of this result M$_w$ has been
fixed and both $\xi\sigma_p$ and the modulation period, T, 
have been considered as free parameters, obtaining 
still the previous value for $\xi\sigma_p$ and T = 
(1.3\(\tiny \begin{array}{l}+0.4\\-0.3\end{array}\)\normalsize) years, 
a period  compatible with a yearly modulation. Then,
both M$_w$ and T have been fixed, while $\xi\sigma_p$ and the 
phase, t$_0$, have been considered as free parameters, 
obtaining the already found $\xi\sigma_p$ value and t$_0$ = 
(140\(\tiny \begin{array}{l}+49\\-43\end{array}\)\normalsize) days, 
a phase compatible with $\simeq 2^{nd}$ June ($\simeq$ 153-th day
in the year 1996). 
The uncertainties on T and t$_0$ are due to the limited
available statistics and to the particular periods of data taking.
The analysis of the maximum likelihood ratio, 
$\lambda = {{\it L}(H_0) \over {\it L}(H_1)}$, 
has been performed to test the goodness of the null hypothesis 
H$_0$ (absence of modulation) with respect to the H$_1$ hypothesis
(presence of modulation according to the given M$_w$ and $\xi\sigma_p$). 
From the definition of the y function we obtain 
$\lambda = e^{[y(H_1)-y(H_0)]/2}$; 
clearly $0 \le \lambda \le 1$.  
A $\lambda$ value close to 1 will imply absence of 
modulation, while a $\lambda$ value close to 0 
will support the presence
of modulation with the given M$_w$ and $\xi\sigma_p$. To perfom a 
quantitative statistical test, one can use the 
variable (-2 ln$\lambda$) 
which is asymptotically distributed as a 
$\chi^2 $ (high values 
of -2 ln$\lambda$ will support presence of modulation). At 90\% C.L. 
the upper tail  $\chi^2_0 $ is equal to 2.7 for 1 degree of freedom;  
in our case -2ln$\lambda$ = 3.14 $> \chi^2_0 $.
Therefore the hypothesis of
no modulation can be rejected in favour of the hypothesis 
of modulation with the given M$_w$ and $\xi\sigma_p$ at 90\% C.L.
At this point the agreement between H$_1$ and the 
experimental data has been analysed by a $\chi^2 $ test
comparing the obtained S$_m$ of table 1 with the expected 
values; a probability of 6\%, to obtain --- only because of 
statistical fluctuations --- a  $\chi^2 $ value higher than we found, 
has been calculated. This probability is mainly limited by 
the S$_m$ values between 8 and 12 keV, which show up to
$\simeq 2 \sigma$ fluctuations from the expected ones. It is, therefore,
evident the relevance of data now under analysis with 
a statistics about 7 times larger than the
present one. We note that here the --- widely considered --- Helm 
SI form factor has been used for iodine \cite{Hel56}; in any case, no 
relevant effects --- within the present errors --- on the quoted
M$_w$ and $\xi\sigma_p$ were observed when adopting different SI
form factors \cite{Lew96}\footnote{To have a qualitative view 
one can overimpose the S$_m$ curve calculated using the found
M$_w$ and $\xi\sigma_p$ values on the S$_m$ of table
1; it is evident that the calculated curve is fully under the 
S$_m$ points and the divergences are quantitatively represented
by the poor value for the C.L.}.

  Moreover, we have also analyzed the NaI(Tl) data with 
the procedure described above, to test the hypothesis of a 
modulation effect due to a spin-dependent (SD) coupled WIMP. 
The calculation has been performed according to \cite{Ber96}, 
but using here for iodine the recently published Ressel SD 
form factor \cite{Res97}. In this way no minimum has been found 
for the y function in the considered mass region.
As mentioned above, 
preliminary results on searches for the annual 
modulation signature have been also obtained 
by using a liquid Xenon scintillator (statistics of
408.2 kgxdays) \cite{Bel96} and the Canfranc NaI(Tl) detectors 
(statistics of 1342.8 kgxdays) \cite{Sar96}. This last experiment 
had a poorer sensitivity than the previous 
one and, to a larger extend, than 
the present one, mainly because of the higher background rate. 
Also these experiments considered only the two extreme periods.
In particular, to obtain a comparison, we have 
examined the Xenon result of \cite{Bel96}. In this case we have 
performed a standard best fit on the S$_m$ values, constraining 
M$_w$ by the accelerator limits and $\xi\sigma_p$ by the more stringent 
results obtained with the NaI(Tl) detector \cite{Ber96}. An indication for 
M$_w  \simeq$ 60 GeV and 
$\xi\sigma_p \simeq 0.5 \cdot 10^{-5}$ pb  --- values compatible 
with those obtained above --- has  been found, but with extremely
poor C.L.; this could be ascribed to the reduced statistics
(a factor $\simeq$ 11 smaller than the present one) and sensitivity
available there. 
 
\begin{figure}[h]
\centerline{\epsfig{figure=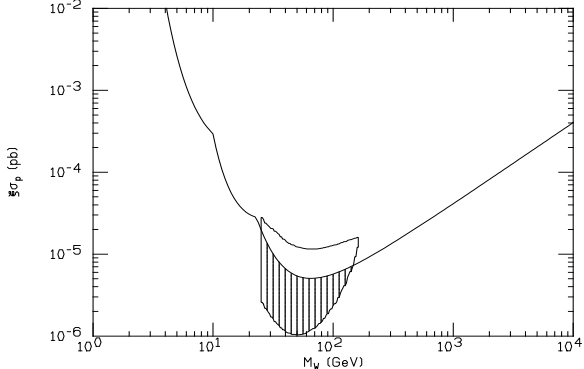,angle=90,height=3.0truecm,width=5.0truecm}}
\caption{$\xi\sigma_p$ versus M$_w$: the region 
allowed by this preliminary analysis at 90\% C.L. 
is shown and superimposed on the best upper limit contour 
for SI interaction, obtained so far [1,13].}
\end{figure}

  In fig. 3 the region allowed at 90\% C.L. --- for a SI 
coupled candidate --- by the obtained $\xi\sigma_p$ and M$_w$ values 
is shown; the best upper limit contour for SI interaction obtained 
so far \cite{Ber96,Reu91} is superimposed. The shaded region represents 
the values for M$_w$ and $\xi\sigma_p$ neither excluded by the present 
analysis nor by the exclusion plot. 
It has been noted that this region
is well embedded in the Minimal Supersymmetric Standard 
Model (MSSM) estimates for neutralino \cite{Bot97,Boa97}.

\section{Is it a robust case?}

  We have always clearly stressed that 
this preliminary result needs further investigations
\cite{Ber97}, however
we are frequently addressed with questions on its "robustness".
We can comment that the present statistical evidence, 
although
in the usual range considered in rare event searches, it is 
not very stringent. Only very large exposure 
would possibly allow to reach a firm conclusion; similar 
exposures will be obtained 
in next future by our experiment, which is continuously running.

A long list of "possible" systematics has been suggested;
however, we have to remark that several of them
--- if present in an experiment --- 
have to be classified as malfunctioning,
not as systematics. This is the case
e.g. of possible uncontrolled energy threshold and PMT gain  
variations.
We stress, in any case, that systematics 
is function of the quality of an
experiment, therefore its nature and level is generally
very different from one experiment to another.
On the other hand, in the annual modulation case 
systematics --- if present --- 
can either simulate the presence of an unexisting 
signal or cancel the presence 
of a real one; therefore, an "a priori" decision 
on the role of its possible "generic" presence
is arbitrary.

In the data taking considered here, we monitor --- in addition 
to the controls by energy calibration -
the external environmental radon,
the HP N$_2$ flux and the
overpressure of the inner Cu box (in which the detectors are),
the temperature and the total and single crystal rates 
over the single photoelectron threshold (i.e. from noise 
to "infinity").
In fig. 4 examples of the behaviours of some of the
parameters in long term are shown.

\begin{figure}[!htbp]
\centerline{\epsfig{figure=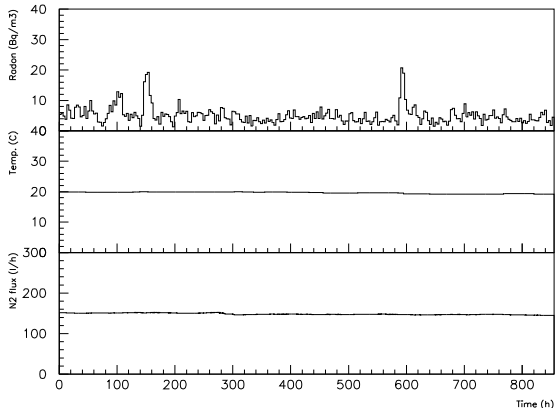,height=4truecm,width=7.5truecm}}
\caption{Example of behaviours of some 
parameters. From top to bottom:
external radon (see text);
operating temperature;
HP N$_2$ flux in the Cu box containing the detectors.}
\end{figure}

In the following, we summarize the main features to control systematics: 
a) data are taken first in winter and, then, in summer time 
(therefore possible positive effects cannot be accounted by isotope decay);
b) the threshold, the PMT gain and electronic line stability 
are verified both by the features of calibration peak and by 
monitoring the rates;
c) the operating temperature is controlled and the environmental
temperature in the installation is not influenced by
external seasonal variations, being conditioned;
d) the stability in the 12-20 keV energy region
allows to verify that no appreciable variation of neutron,
of electromagnetic background
and of environmental conditions were present, although it is not clear 
how all of them could vary
with the same period and phase of a 
possible WIMP signal;
e) the external environmental radon is recorded,
although two levels of sealing in supronyl (maintained in HP N$_2$
long stored underground) isolate the shield containing the Cu box
in which the detectors are closed.
The HP N$_2$ flux in the Cu box 
and its overpressure are monitored;
f) as regards the "possible modulation" spread over 
the crystals, for the sake of completeness 
we recall that the S$_m$ on single crystals 
in \cite{Ber97} have been calculated with another method
(see also the comments there).
When properly addressed, 
observations on spread 
would demonstrate --- as first ---
the necessity of long data taking to avoid 
the difficult position of possible 
"tail" effect (see e.g. also other kinds of rare event searches
when comparing their results from different periods of data taking). 
In any case, we remark that the C.L. found in the overall analysis 
already accounts for the single crystal response (remember
the method features and the ${\it\bf L}$ function definition);
g) the verification of the period and phase of a possible effect 
by the maximum likelihood method 
allows to restrict mainly to systematics effects
able to induce an annual modulation
with 1 year period and $\simeq 2^{nd}$ june phase.
Obviously this will be a stronger constraint when
almost whole years data taking  (periodical calibrations\footnote{such as 
e.g. the time consuming low energy
Compton calibrations needed
to assure in advance the possibility of performing later
also PSA on the collected data \cite{Ber96}.}
and others are obviously needed in any case) will be considered.

Finally, we recall that an almost whole year statistics, in 
improved condition, is already under analysis and the experiment 
is continuously running, therefore in near future 
most of uncertainties will be --- in any case --- overcome.
 
\section{Conclusions}
Considering both the difficulty and the relevance 
of this kind of searches,
a cautious actitude is mandatory (see \cite{Ber97}).
Here we have presented a bare status 
report on first experimental data and analysis method.
We are now further carefully investigating
--- with much larger statistics and in improved
conditions --- the region singled out in fig. 3.

\end{document}